Global regulation of genome duplication in eukaryotes: an overview from the epifluorescence microscope.

John Herrick[1] and Aaron Bensimon[2]

1. e-mail: jhenryherrick@yahoo.fr
   Genomic Vision
   29, rue Faubourg St. Jacques
   75014 Paris

2. e-mail : aaron.bensimon@genomicvision.com
   Genomic Vision
   29, rue Faubourg St. Jacques
   75014 Paris

Abstract

In eukaryotes, DNA replication is initiated along each chromosome at multiple sites called replication origins. Locally, each replication origin is "licensed", or specified, at the end of the M and the beginning of G1 phases of the cell cycle. During S phase when DNA synthesis takes place, origins are activated in stages corresponding to early and late replicating domains. The staged and progressive activation of replication origins reflects the need to maintain a strict balance between the number of active replication forks and the rate at which DNA synthesis proceeds. This suggests that origin densities (frequency of initiation) and replication fork movement (rates of elongation) must be co-regulated in order to guarantee the efficient and complete duplication of each sub-chromosomal domain. Emerging evidence supports this proposal and suggests that the ATM/ATR intra-S phase checkpoint plays an important role in the co-regulation of initiation frequencies and rates of elongation. In the following, we review recent results concerning the mechanisms governing the global regulation of DNA replication and discuss the roles these mechanisms play in maintaining genome stability during both a normal and perturbed S phase.

Introduction: Early single molecule studies of DNA replication

DNA replication has been the focus of study on extended molecules for over 40 years. DNA fibre autoradiography and electron microscopy were the principal technologies used to study the organization of DNA replication on individual molecules spread over a surface. J. Cairns first developed DNA fibre analysis in the 1960s in order to study the replication of the *Escherichia coli* chromosome (Cairns 1963). Later, other researchers employed electron microscopy to image "replication bubbles", or circles of newly replicated DNA formed between un-replicated sequences (Blumenthal 1974). Both techniques provided the first quantitative assessment of replicon sizes and replication fork movement in the metazoan genome.

Huberman and Riggs later applied the method to study DNA replication in mammalian cells (Huberman and Riggs 1966; reviewed in Edenberg and Huberman 1975). These studies formed the basis of the original paradigm concerning the organization of DNA replication in the metazoan genome (reviewed in Berezney 2000). According to the model developed during these studies, the metazoan genome is organized in multiple, tandem units of replication, termed replicons. A replicon is defined as a sequence of DNA that is replicated from a single site, or origin, where DNA synthesis starts, and its size corresponds to the length of DNA replicated from the origin. Following replication initiation, DNA synthesis proceeds either bi-directionally or uni-directionally until advancing replication forks from adjacent replicons merge and replication terminates at random sites. A central tenet of the paradigm involves the organization of replicons into groups, or clusters, of four to ten origins that initiate replication more or less synchronously.

The introduction of fluorescently labelled nucleotides and antibodies along with improved stretching techniques such as molecular combing has resulted in a far more efficient and reliable method for studying genome organization during DNA replication (Jackson and Pombo 1998; Herrick and Bensimon 1999). Fibre fluorography consists of using modified nucleotides such as BrdU, CldU and IdU to label actively replicating sites in the genome. The incorporated nucleotides are then detected on stretched DNA with fluorescently labelled antibodies. Following antibody detection, the labelled DNA is visualized in an epifluorescence microscope as a tandem array of discrete linear signals whose lengths can be directly measured. Initial fluorographic studies confirmed the original autoradiography findings regarding replicon sizes and clustering, and revealed that replicon clusters labelled at the beginning of one S phase were also labelled at the beginning of the following S phase (Jackson and Pombo 1998). Based on these experiments, it was concluded that replicon organization is transmitted and stably maintained from one somatic generation to the next.

Using the fluorographic approach, initial studies on embryonic genome duplication in the *Xenopus laevis in vitro* replication system revealed that replication origins are stochastically and asynchronously activated at intervals of 5 to 20 Kb throughout S phase (Herrick 2000; Blow 2001). Most significantly, the frequency of origin activation was found to increase as S phase advances (Herrick 2000; Marheineke and Hyrien 2001; Herrick 2002; Figure 1A). These unexpected observations contrasted with earlier findings on *Drosophila melanogaster* embryos. In *D. melanogaster* embryos, the data suggest that replication origins are regularly, or periodically, spaced and synchronously activated at the beginning of S phase (Blumenthal 1974). A periodic spacing of replication origins provides a potential solution to what has been called the "random completion problem".

The random completion problem was formulated as an argument against the possibility that replication origins in embryos and other cell systems are either randomly spaced or randomly activated during S phase, because a random distribution of active origins would result in large gaps of un-replicated DNA as cells entered M phase of the cell cycle (Blow 2001). Consequently, cells are expected to undergo "mitotic catastrophe", since un-replicated regions of the chromosomes would subsequently break during mitosis. There are two problems with the assumption that cells will undergo mitotic catastrophe if origins are randomly distributed: 1) it fails to take into account the intra-S phase and G2/M checkpoints in somatic cells that delay mitosis and cell division when DNA remains un-replicated due to replication anomalies; and 2) it overlooks possible mechanisms involving origins that are either continuously "laid down" on un-replicated DNA during S phase, or that fire as "backup" origins when replication forks are impeded or are unable to complete duplication of a replicon (Taylor 1977; reviewed in Gilbert 2007).

It is now widely accepted that the replication program is established prior to S phase and that the timing and order of the replication of sub-chromosomal regions of the genome are stably transmitted from one cell cycle to the next (Ma 1998; Dimitrova and Gilbert 1999). The formation of pre-Replication Complexes (pre-RCs) takes place during late mitosis and early G1-phase (Dimitrova 2002), and pre-RCs are converted to active replication origins in late G1 when CDK levels rise sharply. A pre-determined and reproducible replication program is consistent with the basic assumptions of the random completion problem, but leaves unanswered questions concerning how the cell responds when the replication program is disrupted or delayed by stalled replication forks or by the failure of a pre-specified origin to fire on time (Bechhoefer and Marshall 2007).

In addition to the role of checkpoints, two other solutions to the question of how cells respond to replication anomalies that occur during S phase have been proposed : 1) potential origins of replication exist in excess of the number needed to complete genome duplication, and are activated when replication forks are interrupted (Herrick 2002; Hyrien 2003; Bechhoefer and Marshall 2007); and 2) replication fork rates are coordinated with replicon sizes and can be adjusted dynamically to compensate for an origin if it fails to fire on schedule (Conti 2007, Montagnoli 2007). Emerging evidence indicates that replicating cells rely on both mechanisms, and that these mechanisms are mediated at least in part by the intra-S phase checkpoint response. This review will examine recent fibre fluorography results concerning the regulation of replication origin densities and replication fork rates in a variety of different biological systems. Plausible mechanisms will be discussed that might explain the recent findings concerning the coordinated activation of replication origins and the regulation of replication fork rates during S phase.

The global organisation of replication origins: from early to late replicating domains

The observation that origin density increases as S phase advances was proposed as a potential solution to the difficulties raised by the random completion problem. (Herrick 2000; Herrick 2002; Hyrien 2003). A progressive increase in origin density reflects the increasing probability of activating a replication origin at un-replicated DNA sequences as S phase advances. Consequently, an un-replicated segment of DNA at the end of S phase will be duplicated faster than a similar sized segment at the beginning of S phase because of the greater fork density. The mechanism responsible for increasing the initiation frequency was not, however, explicitly addressed during these studies.

Four possible scenarios can explain the increase in initiation frequency: 1) a continuously synthesized *trans* acting initiation factor accumulates as S phase advances; 2) a progressive change in chromatin structure acts *in cis* to render late firing origins more accessible to replication factors; 3) an initiation cascade, or domino effect, occurs because of a shift in the equilibrium between diffusible replication factors and un-replicated chromatin; and 4) the number of replication forks replicating the genome is maintained at a constant level throughout S phase, and consequently the fork density increases at un-replicated DNA as S phase advances. These scenarios are not mutually exclusive and potentially represent related and overlapping mechanisms.

The existence of a *trans* acting factor that regulates the frequency of origin activation, or origin density, was investigated in experiments using aphidicolin to block entry into S phase in the *X laevis in vitro* system (Marheineke and Hyrien 2001). If such a factor were rate limiting for initiation, its accumulation prior to S phase would be expected to result in a high frequency of origin activation. Nuclei were pre-incubated for 2h in the presence of aphidicolin before the start of DNA synthesis in order to allow any such factor to accumulate. No increase in origin density was observed under these conditions, and it was therefore concluded that a *trans* acting factor does not regulate the frequency of initiation.

Other experiments carried out in parallel indicated that the frequency of initiation is controlled *in cis* (Marheineke and Hyrien 2001). The experiments showed that when replication forks from early activated replicons are blocked by aphidicolin, a caffeine insensitive intra-S phase checkpoint response is induced and activation of late firing origins is prevented. When aphidicolin was removed, origin activation resumed and replication proceeded normally. This demonstrates that the replication program can be stopped and restarted without any significant change in the temporal sequence or density of origin activation (Dimitrova and Gilbert 2000). Thus, the frequency of initiation depends on the amount of replicated DNA rather than the time elapsed since the beginning of S phase.

Additional evidence that the spatio-temporal replication program is fully established and regulated *in cis* was obtained from other fibre fluorography experiments. These experiments showed that the replication program is fixed at two distinct stages of G1 (Li 2003). Nuclei were isolated from mammalian cells at different times in G1 and then incubated in egg extracts from *X laevis* to activate the licensed pre-RCs. One to two hours after mitosis, the replication timing program is set (timing decision point, or TDP). This event determines which chromatin domains will replicate early and which domains will replicate late. Two to five hours after mitosis, a subset of pre-RCs is selected to act as "preferential origins", an event referred to as the origin decision point (ODP).

These results suggest that an excess of potential origins is initially distributed across the genome and then progressively restricted during G1 to specific domains (Figure 2AB). The TDP acts to restrict origins to clusters that will fire synchronously either early or late in S phase; the ODP further restricts the number of origins within clusters that will ultimately serve as active replication origins (Wu and Gilbert 1996). Hence, establishment of the replication program involves the staged suppression of potential replication origins locally (ODP) and globally (TDP) throughout the genome before the onset of S phase (Dimitrova and Gilbert 1999).

The local role of "origin interference" in determining replicon size and plasticity

Fluorographic experiments on human primary keratinocytes (Lebofsky 2006) and the mouse immunoglobulin heavy chain (*igh*) locus (Norio 2005) provided additional evidence *in vivo* for the staged suppression of potential origins in determining replication initiation sites. Using a novel approach called Genomic Morse Code (GMC) to identify a specific region in the genome, the experiments on keratinocytes examined the pattern of origin activation over a 1.5 Mb segment of chromosome 14q11.2. Origins were mapped to "initiation zones" varying in size from 2.6 Kb to 21.6 Kb. Initiation zones correspond to regions in which there is a given probability to initiate DNA synthesis once and only once anywhere inside the zone (Dijkwel and Hamlin 1995a). An origin within a zone is therefore selected stochastically, and all other potential origins within the zone are subsequently suppressed.

In both the experiments on the *igh* locus and keratinocytes, the average distance between initiation zones was found to be only 20 Kb (ranging from 14 to 93 Kb in keratinocytes), which is substantially less than the average replicon size (50 to 300 Kb). The discrepancy can be resolved if each replicon corresponds to more than one potential initiation zone, but only one is activated during a given S phase. The experiments also revealed that the selection of which origin fired in a replicon varied from one S phase to the next (Lebofsky 2006). Therefore, an origin fires stochastically, and origin firing is then suppressed over one to two flanking initiation zones, a phenomenon called "origin interference" (Brewer and Fangman 1993).

Origin interference can be explained by one of two possible mechanisms: origins are either passively replicated and therefore inactivated before they can fire, or they are actively repressed by an unknown mechanism prior to, or immediately after, a preferential origin fires (Brewer and Fangman 1993; Hyrien 2003). These two mechanisms are not mutually exclusive and both potentially play a role in determining replicon size. Although the molecular details of origin interference have yet to be elucidated, the results on keratinocytes and the *igh* locus demonstrate that multiple potential sites of initiation exist within a zone, and multiple potential initiation zones exist within a replicon. (Figure 3AB).

The timing of origin activation at adjacent replicons was also examined. Although the origins of most replicons in a cluster did not fire synchronously, they tended to fire within one to two hours of each other in both experiments (Norio 2005; Lebofsky 2006). Origins in replicon clusters are therefore activated sequentially, which is consistent with a domino effect (Sporbert 2002). In contrast to early embryos where initiation can occur anywhere, it was found that initiation occurs preferentially in inter-genic regions in keratinocytes (Lebofsky 2006), and concomitantly with developmentally regulated changes in chromatin structure and transcriptional activity at the *igh* locus (Norio 2005). In accordance with a hierarchical organisation of sub-foci within foci (Leonhardt 2000), the suppression of potential origins by origin interference establishes a developmentally regulated hierarchy of origin firing, and introduces a considerable degree of redundancy and robustness into the replication program. Together, these results clarify the functioning of origin interference in regulating replicon sizes during S phase and during development.

Additional evidence that replicon sizes are dynamically regulated came from studies that investigated the effect of mitotic re-modelling of replicons and chromatin (Lemaitre 2005). The experiments involved incubating human erythrocyte nuclei in either interphase or mitotic egg extracts from *X laevis*. When incubated in interphase extract, replication in erythrocyte nuclei was inefficient; but when pre-incubated in M phase extract for two hours, the nuclei regained replication competence. Hence, prior mitotic re-modelling of the erythrocyte chromatin was necessary for normal DNA replication in interphase extracts.

Fibre fluorography revealed that re-modelling involved a reduction in the distances between replication origins, and thus a resetting of the replication program by M phase chromatin factors. Incubating erythrocyte nuclei in M phase extract produced a randomization of nuclear attachment sites and a reduction in average chromatin loop size to approximately 20 Kb. It was found that the efficiency of ORC recruitment decreased as loop size increased, indicating that chromatin remodelling influences the number, and possibly the location, of ORC complexes during development. At the same time, origin spacing was observed to decrease from a range of 30 to 230 Kb in untreated erythrocytes to about 25 Kb after treatment with M phase extract. This spacing is similar to that found in sperm chromatin replicating in interphase extract (5 to 20 Kb), and it corresponds closely to the average distance between potential origins observed in the keratinocyte and *igh* experiments (approximately 20 Kb).

The role of replication origin "efficiency" in specifying replicons

The redundancy of replication origins within a replicon suggests that the sites are associated with pre-RCs. Evidence of ORC specified redundant replication origins was provided by two sets of experiments: First, origin densities *in vivo* are normal in human cells expressing low levels of ORC2 (Teer 2006). Second, inter-origin distances in the *X laevis in vitro* replication system are unaltered when the number of nuclei per micro-litre of egg extract is increased (Marheineke and Hyrien 2004). These results indicate that ORC complexes on chromatin exceed the number required to specify and replicate any given replicon (Figure 3AB). These observations complement other results, which show that loading of the MCM2-7 complex is in excess of what is needed for normal replication (Edwards 2002; Hyrien 2003; Woodward 2006), and increases when Cdk2 is blocked (Zhu 2005).

These and other observations suggest a three stage model of replicon specification: 1) one or more ORC complexes bind to a chromatin loop; 2) ORC1 and/or a diffusible replication initiation factor such as CDC7/DBF4 stochastically binds to one of the complexes (Natale 2000); 3) the complex then associates with the nuclear matrix when the origin is activated (Djeliova 2001; Figure 3). The association of replication origins with the nuclear matrix has a long and controversial history (Berezney and Coffey 1975; reviewed in Anachkova 2005), and it remains unclear if this association activates replication origns, or if it has any effect at all on origin firing. Nevertheless, matrix attachment sites have been consistently found near active replication origins (Berezney and Coffey 1975; Anachkova 2005, and references therein), and AT rich sequences strongly correlate with both matrix attachment regions and replication origins (Dijkwel and Hamlin 1995b).

Although the proposed model remains to be verified, the stochastic selection of a replication origin from among multiple potential origins in a sub-chromosomal region is consistent with the events that occur at the time of the origin decision point when ORC1 levels rise (Natale 2000). Recently in the budding yeast *Sccharomyces cerevisiae*, which contains replication origins corresponding to well-defined DNA consensus sequences, replication origins were shown to be selected and activated at random with no apparent pre-determined or reproducible replication program (Czajkowsky 2007). This observation supports a model according to which potential origins are either stochastically specified in G1 or stochastically activated in S phase.

The stochastic activation of replication origins offers one plausible explanation for the relative "inefficiency" of replication origins in eukaryotes (Heichinger 2006). According to the model proposed above, origin inefficiencies can be explained in terms of origin redundancy within individual replicons and the flexibility of origin use from one division cycle to the next, two features that endow the replication program with a considerable degree of robustness. *S. cerevisiae*, for example, contains up to 10,000 ORC binding consensus sequences, but uses only about 400 replication origins during any given S phase (Breier 2004). In *Saccharomyces pombe*, an alternative explanation of origin inefficiency has been proposed according to which a diffusible *trans* acting factor, in this case the *S pombe* Cdc7/Dbf4 kinase, randomly activates replication origins (Patel 2007). Hence, preferential origins might be stochastically specified in G1 and/or randomly activated during S phase. However, additional factors, including chromatin context and epigenetic regulation, likely participate in determining origin efficiencies and the probabilistic firing of replication origins.

Functional coupling of replisomes and sister replication forks

DNA fibre autoradiography experiments originally demonstrated that replication fork rates and replicon sizes are significantly correlated in a variety of organisms including plants (Hand 1975, Kidd 1989). These earlier results have recently been confirmed and extended using the fluorographic approach (Conti 2007). The studies were carried out on mammalian cells and showed that replication fork rates increased in direct proportion to replicon size: smaller replicons tend to be replicated more slowly and larger replicons tend to be replicated more rapidly per unit length. Thus, the time required to duplicate a 1 Mb replicon is similar on average to the time required to duplicate a 100 Kb replicon (approximately 1h), although actual duplication times of individual replicons are expected to vary significantly during a normal S phase (Nakamura 1986; Ermakova 1999; Berezney 2000).

Additional evidence for a functional interaction between replication forks and replication origins within replicon clusters was provided by the observation that replication fork rates at sister forks are co-regulated. Initial autoradiographic results indicated that fork rates within a given replicon do not vary significantly during the period of replicon elongation (Yurov 1979; Berenzey 2000). In contrast to the autoradiographic results, the fluorographic results showed that fork rates can vary up to six fold within an individual replicon as it is being replicated (Conti 2007). Moreover, sister forks in the same replicon changed rates simultaneously, suggesting that deceleration or acceleration did not occur randomly as expected if the change were due to DNA damage or some other non-specific feature of the chromatin. A functional coupling of replisomes is supported by the recent visualization of active replication factories. These studies showed replisomes spatially and temporally couple sister replication fork movement during DNA replication (Kitamura 2006).

These results demonstrate that replication fork rates are coordinated during replicon duplication; but can proceed independently of each other, as indicated by the significant number of unidirectional or asymmetrically moving sister forks (Dubey 1987; Breier 2005; Marheineke 2005; Figure 4). At the molecular level, stalled forks result in the functional uncoupling of the MCM and DNA polymerase activities, followed by hyper-unwinding of DNA and the production of Replication Protein A (RPA) bound ssDNA (Byun 2005; Figure 4B). Since ssDNA is the primary checkpoint signal, functional uncoupling of replisomes in leading to hyper-unwinding corresponds to an early step in checkpoint activation (MacDougall 2007).

Dynamic co-regulation of replication initiation and elongation during S phase

The linear correlation between replication fork rates and replicon sizes either reflects the fact that faster replication forks passively inactivate more potential origins before they can fire, and thus result in correspondingly larger replicons; or it suggests the existence of a homeostatic mechanism that actively coordinates the frequency of initiation and fork progression, with fork progression determining when and where origins fire. Although the former proposal can explain the correlation, it cannot explain why the forks are faster (or slower) in the first place and how they are able to simultaneously adjust their rates either in response to different chromatin environments or as S phase advances (see below).

The latter proposal of a homeostatic mechanism is reasonable in light of the fact that it points to an additional level of control over the kinetics of S phase. What is the evidence that such a mechanism exists? Homeostatic regulation of initiation frequency and fork rates requires that: 1) fork rates respond automatically to changes in initiation frequency that occur during S phase; and 2) origin densities adjust spontaneously to accommodate changes in the fork rates. Such a mechanism can explain the regulated order of origin activation in addition to explaining how the cell might adapt its replication program to unscheduled replication events such as the misfiring of an origin or the interruption of a moving replication fork (Figure 3CD).

*Replication origin densities increase when replication forks are interrupted*

Initial evidence for a homeostatic mechanism that controls replicon size and fork rates during S phase came from the original fibre autoradiography experiments, which demonstrated that in a variety of cell systems new points of replication initiation are recruited when replication forks stall or are blocked because of DNA replication fork inhibitors (Taylor 1977, Francis 1985; Griffiths and Ling 1987; see also Gilbert 2007). The initial investigations showed that inhibiting entry into S phase resulted in a minimal origin spacing of approximately 12 Kb (Taylor 1977; see Gilbert 2007), consistent with the minimal spacing of 14 Kb observed between redundant origins in a single replicon (Lebofsky 2006).

These early observations were later reproduced in another mammalian cell line using the replication fork inhibitor hydroxyurea (HU). HU inactivates the tyrosyl free radical on the R2 subunit of the ribonucleotide reductase enzyme (RNR), thus abolishing its ability to catalyze dNTPs. In that case, inhibiting fork progression led to the activation of an origin that is otherwise inactive during a normal cell cycle: a so called "dormant origin" (Anglana 2003). Hence, replicon size is also under dynamic control during S phase and can adapt spontaneously to changes in replication fork rates.

Although many of the experiments showing a correlation between fork rates and origin density have been carried out in checkpoint or p53 compromised cells, evidence is nevertheless accumulating in support of the proposal that replication origin densities and fork rates are co-regulated. The above observations, for example, are supported by the finding that in yeast cells defective in the ATR homologue Mec1, replication intermediates (RI) in difficult to replicate regions, termed replication slow zones (RSZ), are elevated up to three fold compared to other regions of the yeast genome (Cha and Kleckner 2002). The increase in RI's can be attributed either the activation of dormant origins or to the passive accumulation of replication forks in replication slow zones.

Other fluorographic studies reveal that the correlation between fork rates and origin densities represents a more general phenomenon. Over-production of DNA translesion polymerases beta and kappa, for example, reduces the rate of fork movement while at the same time increasing origin densities (Pillaire 2007). The p53 tumour suppressor gene has been shown to prevent re-replication, and thus its absence might result in an artifactual activation of replication origins that is unrelated to a normal S phase response to genotoxic stress (Vaziri 2003). Importantly, the slowing of replication forks in the DNA translesion polymerase experiments did not induce a checkpoint response although the cells are checkpoint competent. This indicates that the correlation points to a more direct effect of replication fork rates on origin density that occurs independently of checkpoint activation. A more direct effect, however, awaits verification in a system that can tolerate the presence of p53 when fork movement is perturbed.

In agreement with a general phenomenon related to impaired fork progression, replication fork rates in a Bloom's syndrome (BS) cell line are slower than in wild type cells, but the number of new sites of DNA synthesis increases up to 4 fold (Davies 2007; Rao 2007). Treating cells with roscovitine suppressed the excessive origin firing, indicating that activation of dormant origins in BS cells depends on S phase kinases (SPK) such as Cdk2 and Cdc7/Dbf4 (Davies 2007). Similar findings that altered replication fork rates result in increased origin densities have been reported in a yeast replication mutant defective in the Claspin homologue Mrc1 (Tourriere 2005). These experiments showed that slower forks correlate with higher origin densities in the presence of HU but not in its absence. This observation is consistent with the firing of late origins when fork movement is perturbed (Tourriere 2005), and supports the proposal that RI accumulation in replication slow zones in Mec1 compromised cells corresponds to "dormant origin" activation rather than passive accumulation of replication forks. In contrast to these results, a Werner Syndrome cell line bearing a mutation in the WRN helicase gene did not reveal a relationship between perturbed fork rates and origin density, since these cells exhibit perturbed fork movement but normal origin spacing (Rodríguez-López 2002). Analysis of a mutant of the yeast WRN homologue, Sgs1, also revealed normal origin spacing, but showed that fork rates increase in this genetic background (Versini 2003).

A potential explanation for the increase in fork density in response to perturbed elongation was provided by experiments using the *X laevis* replication system. These experiments revealed that loading of excess MCM/CDC45 complexes onto chromatin plays a potentially important role in dormant origin activation when replication forks are blocked (Woodward 2006). A functional role for the observed activation of dormant origins *in vivo* was revealed by other studies on oncogene induced cellular senescence (OIS). During tumorigenesis, cellular senescence induced by Ras oncogene activation restrains cell proliferation and transformation (Di Micco 2006). Ras oncogene expression produces reactive oxygen species (ROS) and results in a hyper-proliferative phase that induces the DNA damage response (DDR) followed by OIS. If the damage response is abrogated experimentally, cells continue to proliferate and transformation occurs. It was found that oncogene activation of the DDR depends on DNA replication and induces a period of hyper-replication (HR) during which origin densities increase nearly two fold. This observation is consistent with earlier studies, which showed that overproduction of another oncogene, CMYC, caused locus specific hyper-replication and amplification of the RNR R2 gene (Kuschak 2002). In agreement with the findings on Ras, MYC over-expression elicits ROS accumulation (Matsumura 2004) and increased origin activation (Dominguez-Sola 2007).

*Replication fork rates increase when origins of replication are inactivated.*

Additional evidence for the homeostatic regulation of initiation frequency and fork rates came from early autoradiographic studies that showed inhibiting replication origins increased replication fork rates. In plants an increase in replicon size induced by the plant hormone trigonelline resulted in a 1.6 fold increase in fork rates (Mazzuca 2000). The same phenomenon was observed in a hamster cell line, ts BN2, which is temperature sensitive for replication initiation (Eilen 1980). After a shift to non-permissive temperature, the interval between adjacent initiation sites was found to increase and the corresponding frequency of initiation events decreased. At the same time, replication fork rates increased by 30 %. Although it remains to be established if the increase in replication fork rate in the ts BN2 mutant directly depends on the decrease in origin density, the correlation is nevertheless consistent with the emerging evidence that origin densities and replication fork rates must be carefully balanced and thus co-regulated during S phase.

The cell line used in these experiments, ts BN2, contains a temperature sensitive mutation in the RCC1 (Regulator of Chromosome Condensation 1) gene. RCC1 is implicated in coupling S phase to G2/M; and it exerts its effect via the Ras-related RanGTPase, which is a regulator of nuclear transport and Cdc2/cyclinB activation (Moore and Blobel 1993; Clarke 1995; Takizawa 1999). Consistent with its role in S phase progression, expression of the RCC1 gene is also up-regulated by c-myc (Tsuneoka 1997), and RCC1 and Ran have been proposed to play a direct role in the assembly and/or activation of pre-RCs (Hughes 1998). Hence, the nuclear import of replication factors might play an important role in regulating genome duplication and the kinetics of S phase (Cox 1992; Ellis 1997). Although it remains to be established if the increase in replication fork rate in the ts BN2 mutant directly depends on the decrease in origin density, the correlation is nevertheless consistent with the emerging evidence that origin densities and replication fork rates must be carefully balanced, and thus co-regulated during S phase,

More recently, the dynamic correlation between fork rate and replicon size was confirmed during a number of experiments that investigated inhibitors of CDK proteins in mammalian cells. In one study a small molecule inhibitor of the CDC7 kinase was used to study its effects on DNA replication (Montagnoli 2007 submitted). Inhibiting CDC7/Dbf4 was shown to block phosphorylation of MCM2 (Tenca 2007, Montagnoli 2007 submitted), and at the same time resulted in a corresponding increase in the distances between replication origins (Montagnoli 2007 submitted). Hence, replication initiation was compromised by the compound, but the effect of inhibiting initiation was compensated by a proportional increase in the replication fork velocity.

These observations are supported by similar findings in yeast, in which two separate studies demonstrated that inhibiting initiation stimulates fork rates (Shimada 2002; Semple 2006). The experiments involved either depleting Orc6 or inactivating Orc2 in late G1, which reduced the efficiency of replication initiation. Approximately half the number of origins was activated in Orc6 depleted cells, but the lower origin density was simultaneously compensated by a corresponding two fold increase in fork rates (Semple 2006). This effect was attributed to the fact that cells with fewer origins firing have more nucleotides available for elongation.

The same effect occurs in bacteria under certain conditions, suggesting that it reflects a general principle of DNA replication and DNA metabolism rather than a specific adaptation of any particular cell system (reviewed in Herrick and Sclavi 2007). Together, these results suggest a functional link between nucleotide pool size and replication fork density, and indicate that the two must be carefully balanced at replication forks in order to maintain a restricted range of fork rates at all replicating sites.

Global regulation of replication fork rates as S phase advances

Early autoradiography studies demonstrated that replication fork rates increase up to three fold toward the end of S phase when heterochromatin is replicated (Houseman and Huberman 1975). More recent fibre autoradiography results from human cells confirm and extend the earlier findings of faster fork rates as S phase advances and identified at the same time a mid-S phase replication slow zone termed the 3C pause at a R/G band border (Takebayashi 2001). These studies also correlated fork rates with changes in replication foci patterns as the transition from replicating euchromatin to replicating heterochromatin took place. Replication forks advanced at a rate of 1.2 Kb/min during early S phase and slowed down at mid-S phase to 0.74 Kb/min. In contrast, during late S phase fork rates steadily increased to a maximum of 2.3 Kb/min. The slowest rate of replication fork progression occurred at the R/G boundary while the fastest rate occurred when heterochromatin was being replicated. These results therefore agree with the earlier autoradiographic results concerning an increase in fork rates as S phase advances (Housman and Huberman 1975; see also Dimitrova and Gilbert 1999), and suggest that features of the chromatin modulate fork rates during the transition from early to late replicating domains.

Numerous reports have shown that DNA replication fork rates depend directly on dNTP levels *in vivo* and *in vitro* (Stano 2005). The enzyme ribonucleotide reductase is universally responsible for dNTP synthesis (Nordlund and Reichard 2006). In eukaryotes, production of the enzyme is induced at the G1/S phase transition and is elevated in response to DNA damage. In early S phase, dNTP concentrations *in vivo* are low and increase continuously until they reach a maximum at the end of S phase (Walters 1973). Since replication fork rates increase three fold toward the end of S-phase, dNTP synthesis appears to have a direct effect on fork rates. Consistent with such an effect, fluorographic analysis has shown that the supply of exogenous dNTPs accelerates replication fork speeds by up to three fold in early but not late S-phase (Malinsky 2001). Consequently, fork rates in early S phase are limited by dNTP availability, indicating that dNTP levels globally regulate replication fork rates during S phase.

Involvement of the ATM/ATR pathway in the global and local regulation of replication origins

When DNA damage occurs DNA synthesis is arrested and cell cycle progression is prevented (Feijoo 2001). The ATR/ATM intra-S phase checkpoint response is activated under these conditions and the effector kinase Chk1 phosphorylates and inhibits Cdc25C, thus preventing activation of the Cdc2/cyclinB complex and entry into mitosis (Sanchez 1997). Early observations that DNA replication is required for checkpoint activation and that Chk1 is induced at G1/S in unperturbed cells suggested that the checkpoint also participates in a normal S phase (Kaneko 1999; Lupardus 2002).

Several studies have demonstrated that late origin firing is regulated by the ATR/ATM pathways via the downstream targets Chk1, Cdc25A, and Cdk2 in the absence of DNA damage (Dimitrova and Gilbert 2000; Sorensen 2004). Initial evidence for a Chk1 dependent pathway that regulates replication origins in the absence of DNA damage came from experiments that involved specifically inhibiting Chk1 activity during a normal S phase in human fibroblasts (Miao 2003). These studies revealed that late firing origins become early firing when Chk1 activity is abrogated by a variety of inhibitors including caffeine. The shift from late to early firing resulted in an increased initiation frequency in early S phase.

Later studies confirmed these observations. Abrogation of Chk1 was associated with a transient stimulation of Cdk activity, increased initiation of DNA replication, massive induction of ssDNA and breakage of DNA (Syljuasen 2005). Inhibiting initiation in Chk1- cells by down-regulating Cdk2, Cdc45 or by treatment with the SPK inhibitor roscovitine reversed the genetic instability observed in the Chk1- cells (Syljuasen 2005). Fibre fluorography experiments further demonstrated that Chk1 abrogation, but not Chk2, results in a two fold increase in origin firing at the beginning of S phase and shorter inter-origin distances. Concomitantly, replication fork rates are reduced by 25 to 50 percent, indicating that fork rates decrease in response to increases in origin density (Marheineke and Hyrien 2004; Petermann 2006). An increase in homologous recombination events was ruled out as the cause of the observed slower fork rates (Petermann 2006), suggesting that some other factor becomes limiting for elongation under these conditions. Thus, Chk1 regulates origin densities in actively replicating regions of the genome during S phase (Maya-Mendoza 2007)

The regulation of late origin firing appears to be controlled by a simple feedback mechanism that communicates between early and late firing replication origins (Shechter 2004). Initiation of DNA replication results in single strand DNA (ssDNA) at replication origins. The RPA protein binds ssDNA and activates the ATM/ATR checkpoint response. This in turn results in origin interference (OI) and a block to late origin firing. ATR principally governs Chk1 activity (Kaneko 1999), and it mediates origin interference by inhibiting Cdc7 at late firing origins, while ATM mediates origin interference by inhibiting the Cdk2 kinase. When DNA at early firing origins is replicated the formation of dsDNA relieves origin interference and late origins subsequently fire, resulting in the progressive and orderly activation of origins throughout S phase (Shechter 2004).

In addition to its role in regulating late origin firing, the ATR/ATM pathway also regulates the activity of ribonucleotide reductase. The transcription regulator Rfx1, a homologue of the yeast transcription factor Crt1, represses RNR-R2 gene transcription. During DNA damage, RNR gene transcription is up-regulated in an ATR/ATM dependent manner (Nordlund and Reichard (2006). Recent results indicate that Chk1 directly mediates RNR up-regulation, and up-regulation is both Rfx1 dependent and independent (Lubelsky 2005). Since Chk1 activation leads to degradation of Cdc25A and persistent inhibition of Cdk2 (Mailand 2000), this suggests that Chk1 coordinates RNR activity and SPK regulated origin activation (Figure 5).

In yeast, the Mec1/Rad53/Dun1 pathway plays a similar role in inducing RNR production by relieving Crt1 repression of the RNR2 promoter. It has been suggested that Rad53 plays two principal roles during S phase: 1) it inhibits late origin firing in a checkpoint dependent manner (Santocanale 1998), and 2) it up-regulates RNR in response to DNA damage (Huang 1998). Like Chk1, abrogation of Rad53/Cds1 also results in late origins firing earlier in a normal S phase (Shirahige 1998; Hayashi 2007). It remains to be directly shown, however, if these proteins participate in the regulation of RNR during a normal S phase. Nevertheless, these observations suggest that Chk1, Rad53 and Cds1 participate in coordinating dNTP pool sizes with replication fork densities during both a normal and perturbed cell cycle (Figure 5). Such a mechanism provides an appealing explanation of the repeatedly observed correlation between replication fork rates and origin densities.

The role of late replicating domains in the global regulation of genome duplication

In somatic cells, the genome is divided between euchromatin domains, which tend to replicate early, and heterochromatin domains, which tend to replicate late. In contrast to somatic cells, a clear distinction between euchromatin and heterochromatin does not apply to *X laevis* embryos, and the same pattern of intra-nuclear replication foci persists throughout S phase (Mills 1989; Figure 1 CD). Nevertheless, two distinct replication regimes appear to exist in *X laevis* egg extracts despite the different chromatin organization in this system. In egg extracts, the two regimes are distinguished by an abrupt transition (break point) to higher origin densities in the second half of S phase (Figure 1A). This suggests that features other than chromatin organization alone are important in determining replication kinetics during the eukaryotic S phase.

The observations that slowing replication forks results in an increase in origin densities under a variety of conditions indicates a potential role for replication fork slow zones in increasing the activation of replication origins in late S phase (Figure 1B). Common fragile sites and replication fork slow zones are often associated with the boundaries between early and late replicating chromosome bands (Glover 2006; Debatisse 2006). In accordance with the increased numbers of replication intermediates found in yeast replication slow zones (Cha and Kleckner 2002), the presence of RSZs at the interface between early and late replicating chromatin is one possible explanation for the break point observed in egg extracts.

What is the evidence that origin densities also increase during late replication in somatic cells? The transition between the two replication regimes in egg extracts occurs after approximately 50 percent of the genome has been duplicated (see Figure 1A). In somatic cells, the transition between R/G bands likewise occurs after approximately 50 % of the genome has been duplicated (Takebayashi 2001). In yeast, Cdc45 associates with early origins in G1 just before DNA synthesis; but only associates with late origins in S phase, again after approximately 50 % of the genome has been duplicated (Aparicio 1999). This suggests that changes in Cdc45 and SPK activities coincide with the abrupt increase in origin activation in late S phase after approximately half the genome has been duplicated.

Other studies on CHO cells suggest that Cdc45 recruits Cdk2 to replication foci resulting in histone H1 phosphorylation and an extensive chromatin de-condensation that correlates with active DNA synthesis (Alexandrow and Hamlin 2005). H1 phosphorylation is lowest in G1, increases significantly and reaches a maximum by G2/M. In cycling *Xenopus* egg extracts, an abrupt increase in Cdc2/cyclinB driven H1 phosphorylation occurs at the end of S phase and apparently coincides with the increase in origin density. Cdc2 and Cdk2 activities have been reported to have overlapping roles in activating replication origins (Aleem 2005). Consequently, the abrupt increase in H1 phosphorylation toward the end of S phase in higher eukaryotes could reflect an accelerated transition between replicated and un-replicated chromatin that is due to an elevated frequency of replication initiation driven by Cdk2/Cdc2.

Additional evidence that origin densities increase as S phase advances comes from the original fluorescence studies on replication foci (Nakamura 1986; Manders 1992). In late S phase fluorescently stained regions of replication associated with perinucleolar heterochromatin are larger and more intense than replicating euchromatin in early S phase. It was therefore concluded that there must be more numerous replicon clusters, and thus higher origin densities, in each fluorescent region corresponding to late replicating DNA (Manders 1992). Based on the observations that potential origins are spaced approximately every 15 to 20 Kb in somatic cells, origin densities in inactive late replicating chromatin would be expected to be higher than origin densities in early replicating euchromatin, with origin spacing ranging from 15 to 95 Kb versus 50 to 300 Kb. This prediction remains, however, to be verified.

The predicted increase in origin density as S phase advances would be expected to coincide with the observed increase in dNTP levels and fork rates in somatic cells. In contrast to somatic cells, replication fork velocity has been reported to decrease rather than increase as S phase advances in *Xenopus* egg extracts (Marheineke and Hyrien 2001). This discrepancy might be specific to the *in vitro* system, or it might reflect the absence of RNR gene expression and thus exhaustion of dNTPs in late S phase. In somatic cells, an increasing fork density would be expected to stimulate the checkpoint response and up-regulate RNR activity, which can explain the observed increase in replication fork rates at the end of S phase. Thus, levels of RNR activity, which are maximal at the end of S phase (Malinski 2006), play a potentially important role in determining the overall replication kinetics of early and late replicating chromatin, and hence in differentiating between euchromatin and heterochromatin.

Conclusions and perspectives

The sequential origin activation within replication foci and the transition between early and late replication regimes reviewed here can be characterized as an initiation cascade, or domino effect, that operates simultaneously at three levels (Herrick 2000; Leonhardt 2000; Sporbert 2002): 1) disassembly into a nucleoplasmic pool of rapidly diffusing replication factors, such as Cdc7/Dbf4 and PCNA (Sporbert 2002), and reassembly at un-replicated DNA results in increasingly higher origin densities at the end of S phase (Herrick 2000; Marheineke and Hyrien 2001); 2) cumulative chromatin de-condensation promotes access of replication factors to replication origins as S phase advances (Manders 1996; Alexandrow and Hamlin 2005); and 3) Chk1 inhibition of late firing origins is relieved after approximately half the genome is duplicated (Miao 2003; Syljuasen 2005), an event that presumably coincides with enhanced CDC45 chromatin association at mid-S phase (Aparicio 1999).

The progressive activation of replication origins and foci raises important questions concerning how the mechanism operates. One proposal suggests that termination of replication in one focus specifically triggers replication in a neighbouring focus, implying a mechanism that coordinates the sequential activation of replication origins and replication foci across the genome (Sporbert 2000). Initiation of DNA replication at new sites in mammalian cells, for example, coincides with the disassembly of replication factories at early sites (Dimitrova and Gilbert 2000). Late firing origins, however, can be activated without completion of early replication (Dimitrova and Gilbert 2000), indicating that origins can fire independently of each other as is the case in yeast (Patel 2006; Czajkowsky 2007) Alternatively, replication is not coordinated by a specific mechanism, but instead is randomly activated with increasing probability as a function of the amount of DNA replicated locally at adjacent foci and globally throughout the genome (Herrick 2002; Patel 2006; Czajkowsky 2007). Given the licensing of replication before S phase in an unperturbed cell cycle, it is likely that these proposals are not mutually exclusive; and the actual replication program combines features of both, with origins being non-randomly specified *in cis* but randomly activated *in trans*.

The role of the checkpoint in globally regulating origin activation and genome stability is not as clear as the roles of positively acting factors such as components of the licensing system and the SPKs. Replication slow zones play a potential role as *cis* acting signals that differentiate between sub-chromosomal replication domains (Debatisse 2006); and deregulation of replication origins might contribute to the intrinsic instability of fragile sites, since over-initiation results in replisome collision and DNA fragmentation (Davidson 2006). Currently, it is unknown how widespread replication slow zones are in different eukaryotic genomes, or if their non-random locations consistently coincide with R/G boundaries. Nevertheless, replication fork slow zones and fragile sites have been proposed to up-regulate the checkpoint in order to delay mitosis until complete duplication of the genome (Cha and Kleckner 2002; Debatisse 2006). Recent findings cast doubt on such proposals, since completion of replication is not under checkpoint surveillance (Torres-Rossel 2007). Other reports, however, indicate that the checkpoint is more strongly enforced at late firing origins during S phase, explaining in part their delayed activation (Seiler 2007). In fission yeast, late firing origins appear to be more efficient than early firing origins, where efficiency refers to the frequency of their use rather than the timing of their activation (Eshaghi 2007). The efficient origin activation in late S phase implies that most available origins, rather than a subset of preferential origins, are activated at that time.

The activation of dormant origins in response to retarded replication fork movement is somewhat paradoxical, since stalled replication forks are responsible for invoking the checkpoint and blocking origin firing (Merrick 2004). Based on the original fibre autoradiography experiments, the activation of dormant origins was predicted to occur primarily in regions of the genome already undergoing DNA replication (Griffiths and Ling 1984), and this prediction has recently been verified (Maya-Mendoza 2007). The increased origin density observed in replicating regions either during genotoxic stress or during checkpoint inhibition suggests a functional interaction between Chk1 and Cdc45/MCM in locally regulating origin densities (Figure 2CD).

Two lines of evidence support such an interaction: levels of chromatin associated CDC45 are reduced when Chk1 is activated in cells exposed to low levels of DNA damaging agents (Liu 2006; Heffernan 2007); and enhanced loading of CDC45/MCM onto chromatin occurs when Cdk2 function is compromised (Zhu 2005). In addition, it has been shown that high levels of genotoxic stress stimulate Chk1 degradation, thus down-regulating its activity (Zhang 2005). Hence, degradation of Chk1 during exposure to DNA damaging agents might result in dormant origin activation locally in replicating regions.

Other factors likely play a role in the enhanced loading of Cdc45. MYC expression, for example, can over-ride the checkpoint and result in hyper-replication (Kuschak 2002). Hyper-replication appears to be due to MYC activation of CDK2 (Li and Dang 1999), and recent results support a more direct role for MYC in activating replication origins (Dominguez-Sola 2007). A functional interaction between MYC, Cdc45, SPKs and/or Chk1 in regulating dormant origins remains, however, to be demonstrated.

In metazoans, the second half of S phase is accompanied by a steady increase in dNTP pools, but how RNR levels are coordinated with origin densities remains unclear. Does the apparent up-regulation of RNR at mid-S phase result in the subsequent down-regulation of the checkpoint in late S phase? Down-regulation of the checkpoint in response to increasing dNTP pools can be explained by the conversion of ssDNA to dsDNA and rapid nascent strand fusion into bulk DNA (Figure 5B). If so, the actual signal, whether a nucleotide, dsDNA or some other factor, and the transducing agents that execute the down-regulation are largely unknown. Nevertheless, RNR is also up-regulated during translesion DNA synthesis, which facilitates DNA replication through damaged DNA (Huang 1998; Nordlund and Reichard 2006). Enhanced replication through DNA lesions in turn attenuates checkpoint signalling (Barkley 2007), and thereby allows dormant and late origins to fire.

How does the cell impose and maintain a distinct late replicating regime? One possibility is the coupled activation and down-regulation of Chk1, which occurs locally in replicating regions encountering high levels of genotoxic stress (Zhang 2005), and which might occur globally during replication through R/G boundaries (Figure 5B). Chk1 activation during replication stress results in its degradation (Zhang 2005; Mamely 2006; Gewurz and Harper 2006 and references therein), which would therefore allow replication to resume via dormant origin activation even if forks are irreversibly blocked. The reported activation of replication origins during replication restart is consistent with the proposal that dormant origin activation is related to checkpoint attenuation rather than to checkpoint activation (see Grossi 2007). Checkpoint attenuation initiated by RNR/polκ/polβ up-regulation and enforced by Chk1 degradation could therefore stably impose on the cell a late replicating regime that is characterized by decreased origin distances and increased fork rates (see Seiler 2007). The switch from a checkpoint-activated replication regime to a checkpoint-attenuated regime might therefore coincide with an enhanced loading of Cdc45/MCM proteins in late S phase and the transition from S to G2 and M phases (see Figure 2).

Based on the observations reviewed above the following model can be proposed: 1) replication initiation at early firing origins activates Chk1 imposed origin interference at late firing origins (Miao 2003; Shecter 2004); 2) replication through replication slow zones, for example, stimulates the checkpoint and up-regulates RNR and translesion DNA polymerases at mid and late S phase (Huang 1998; Pillaire (2007); 3) the checkpoint is subsequently down-regulated in response, and origin interference is relaxed at late firing origins (Zhang 2005; Mamely 2006); 4) origin density and fork rates consequently increase and Cdc2/cyclinB is activated (Kramer 2004; Niida 2005), which signals the transition to G2/M. Although each of these points remains to be verified, preliminary evidence exists in support of them, including the checkpoint-independent activation of dormant origins when translesion DNA polymerases are over-expressed (Pillaire 2007), as well as the finding that active Chk1 antagonizes replication fork movement (Seiler 2007).

Why is there late replicating DNA? Early replicating DNA is characterized by gene rich euchromatin while late replicating DNA is characterized by gene poor heterochromatin (Klevecz and Keniston 1975; Holmquist 1982). Euchromatin is replicated slowly and its replication invokes the checkpoint response that delays replication of heterochromatin (Kaneko 1999; Dimitrova and Gilbert 2000; Lupardus 2002; Marheineke and Hyrien 2004; Shecter 2004). Heterochromatin, in contrast, is replicated rapidly, and the apparently high frequency of initiation implies a down-regulation of the checkpoint response and a relaxation of origin inhibition. Down-regulation of Chk1 is associated not only with increased initiation frequency (Miao 2003; Syljuasen 2005), but also with activation of Cdc2-cyclin B and entry into G2/M (Kramer 2004; Niida 2005). Imposition of a strong block to late firing origins might therefore act to guarantee the complete duplication of early replicating sub-chromosomal domains, and hence the duplication of all *genes* before mitosis begins. This suggests that the late replication of heterochromatin acts as a buffer against premature entry of un-replicated euchromatin into G2/M, thus protecting the integrity of the genome and maintaining its stability until all genes have been successfully duplicated. The biased accumulation near heterochromatin of repetitive DNA, a marker of genetic instability, indicates that late replicating DNA confers an adaptive advantage on the cell by obviating DNA breakage in gene rich regions during mitosis (LeBeau MM 1998). Rather than selfish or junk DNA, heterochromatin, in foregoing early replication, is perhaps better characterized as "sacrificial" DNA.

Acknowledgements: The authors would like to thank John Bechhoefer for comments and careful reading of the manuscript, and Bianca Sclavi, Christophe Henry, Pierre Walrafen, Jean-Pascal Capp and Jun Komatsu for helpful discussions. The authors would also like to thank the three anonymous reviewers whose comments significantly improved the manuscript.

Captions

Figure 1 : A) Nucleation density vs. fraction of replicated DNA. *Xenopus laevis* sperm chromatin was incubated in *X laevis* egg extracts resulting in one complete round of genome duplication. Replication eyes were visualized on linearized DNA molecules and distances between the centers of adjacent eyes were measured. These measurements allow for an assessment of the initiation frequency (based on fork density) as a function of replicated DNA. Two distinct replication regimes are observed: at 5 to 50 % replication, fork densities increase two fold; at 50 to 90 % replication, fork densities increase up to twelve fold. The break point at 50 % suggests an abrupt transition in replication kinetics as S phase advances. B) Diagram of the initiation cascade, or domino effect. Replication factors assemble from the nucleoplasmic pool at an origin inside an early replication focus. Although replication origins are clustered, initiation is not simultaneous but occurs sequentially within a relatively short window of time. This results in modest but significant correlations between adjacent origin activation times (Blow 2001; Jun 2002). Disassembly of initiation factors is followed by reassembly from the nucleoplasmic pool at adjacent origins. Sequential disassembly at replicated DNA and reassembly at un-replicated DNA results in an initiation cascade, or domino effect, according to which activation of one origin increases the probability of activation of adjacent origins in a cluster (Herrick 2000; Sporbert 2002). Similarly, termination of replication at an early focus results in the disassembly of replication factors followed by reassembly at a neighboring later replicating focus. Consequently, as DNA is replicated, the equilibrium between replication factors in the nucleoplasmic pool and un-replicated DNA shifts in favour of increasing origin densities. C) Diagram showing replication foci distributions during S phase in *Xenopus laevis* egg extracts. D). Diagram

(adapted from Leonhardt 2000) showing morphological changes in replication foci distributions between early, mid and late S phase. Black dots: replication foci; grey region: nucleolus. Recent studies suggest that replication slow zones (RSZ) mediate the transition between early and late S phase, providing a hypothetical explanation for the observed break point in *X. laevis* origin activation (indicated by "?"). NOTE: The alignment of the figures is for illustrative purposes in order to compare and contrast general organizational features of S phase in both embryonic and somatic cells. It is not being suggested that the replication programs in these different systems are identical (see text for discussion).

Figure 2: A possible mechanism for the homeostatic coordination of fork rates and replicon sizes. A) At the TDP, early (red circles) and late (black circles) replication origins are specified (adapted from Li 2006). B) At the ODP, a replication origin is selected to designate a replicon (green circle). C) Origin interference (OI), represented by Chk1, inactivates adjacent potential origins and displaces replication initiation and elongation factors from the potential origins. Consequently, more dNTPs will be available for replication forks belonging to the replicon. D) The elongation factors, represented by Cdc45, are redistributed and recruited from the inactive potential origins to the preferential origin. Since larger replicons correspond to more potential origins, a correspondingly larger amount of replication factors will be redistributed within a focus and locally recruited to the active origin that specifies the replicon. Hence, large replicons will tend to be replicated proportionally faster than smaller replicons if each focus contains an equivalent amount of replication factors. Likewise, inhibiting CDKs results in larger replicons and the concomitant recruitment of CDC45/MCM and possibly RNR to chromatin (Edwards 2002; Zhu 2005; Woodward 2006), which can explain the corresponding increase in replication fork rates under these conditions. Conversely, abrogating OI will result in excess origin firing and proportionally slower forks.

Figure 3: Random selection of initiation zones (IZ) within a replicon (adapted from Anachkova 2005). A) One of three ORC/IZ complexes (colored circles 2 and 5) associated with a single chromatin loop randomly attaches to the nuclear matrix during one cell division cycle. B) During a second cycle, after chromatin remodelling, a different ORC/IZ complex specifies the same replicon (coloured circles 3 and 6). Each replicon (brackets) corresponds to two or more initiation zones (IZ). Once a potential origin fires, the replicon is specified by a locally acting origin interference (OI) mechanism. C) In a perturbed S phase, blocking a replication fork results in the local activation of an inefficient, or dormant, origin (ORC unattached to the nuclear matrix) and a reduction in replicon size (HR). Two possible mechanisms involving CDC45/MCM loading correspond to either the relaxation of OI within a cluster (see text), or the activation of dormant origins that otherwise would be passively replicated in the absence of stalled forks. D) Conversely, inactivation of initiation factors such as the SPKs, or the failure of an origin to fire, results in a larger replicon size and compensating faster replication fork rates due to the recruitment of extra replication elongation factors such as RNR, Cdc45 and MCM proteins (small circles; Edwards 2002, Zhu 2005, Woodward 2006, Montagnoli 2007).

Figure 4: A) Replisome coupling at sister replication forks. Replication forks are simultaneously processed by coupled replisomes as DNA is spooled through replisomes during unperturbed DNA synthesis. Coupling allows for the spontaneous readjustment of fork rates in response to changes in chromatin structure or encounters with other proteins such as RNA polymerases. Large arrows: direction of spooling through the replisome. B) When a DNA lesion is encountered, the MCM helicase complex (yellow) and DNA polymerase complex (pink) disassociate (Byun 2005), which results in an uncoupling of replisomes and subsequent asymmetric fork progression (red crosses). The DNA becomes hyper-unwound and triggers the checkpoint response when ATR-ATRIP (not shown) binds RPA (purple circles) associated with ssDNA, thus initiating the checkpoint cascade. The Claspin complex (large brown circles) stabilizes the uncoupled complex and prevents fork collapse. CAF-1 and Asf1 are shown in green. Histones are depicted as small brown circles. Other factors involved in stabilizing replication forks (eg. checkpoint proteins and helicases such as WRN, BLM and RRM3) have been omitted. The diagram is not drawn to scale and protein orientation is represented arbitrarily for illustrative purposes.

Figure 5: Simplified diagrams of the checkpoint mediated pathway in the metazoan replication initiation/elongation cycle. Three different modes of the checkpoint response can be distinguished: 1) during a normal S phase, Chk1, but not Chk2, inhibits late firing origins in response to early origin firing; 2) when replication fork movement is moderately perturbed, for example by low doses of DNA damaging agents, Chk1 origin inhibition is locally abrogated in replicating regions and dormant origins fire in a checkpoint independent manner; and 3) when double strand DNA breaks occur at high doses of DNA damaging agents, Chk1 inhibits both replication initiation and elongation and Chk2 initiates replicative senescence. A) During an unperturbed cell cycle, the E2F1-3/pRB/S-CDK pathway effects the transition between G1 and S phases. The ssDNA/RPA complex activates a low level checkpoint response that is both Cdc25A/CDK2 dependent and independent (Sorensen 2004; Liu 2006; Heffernan 2007). The checkpoint response down-regulates Cdk2 and Cdc7/Dbf4 (here represented by S-CDK) through feedback from ssDNA formed at newly initiated replication origins (Schecter 2004). Cdc45/MCM proteins are in turn down-regulated at dormant origins and local imposition of origin interference (OI) occurs. Activation of the checkpoint results in Rfx1 down-regulation and the graduated induction of RNR activity as S phase proceeds

(Huang 1998; Lubelsky 2005), thus effecting the transition between initiation and elongation. Elongation results in nascent strand fusion into bulk chromatin and subsequent down-regulation of the checkpoint. Mechanisms of checkpoint recovery have been omitted, and are subsumed under dsDNA. Red arrows indicate the principal pathway of the events leading from initiation to elongation during physiological S phase. B) In the absence of a strong DDR either due to checkpoint abrogation or to a delay in checkpoint activation, negative feedback switches to positive feedback and origin activation (red arrows). Oncogene over-expression (Ras/Myc/Jun) potentially over-rides the normal G1/S phase checkpoints and origin interference (Leone 1997; Clark 2000; Kuschak 2002; Maclaren 2003; DiMicco 2007, Dominguez-Sola 2007), thus resulting in checkpoint independent hyper-replication (HR) and rapid fusion of ssDNA into bulk chromatin (Woodward 2006). External DNA damage (lightning bolt) likewise results in the uncoupling of the DNA polymerase complex from the helicase complex, followed by hyper-unwinding and the exposure of ssDNA (Byun 2005). Preceding an amplified checkpoint response, Cdc45/MCM activate dormant origins and HR occurs (Edwards 2002; Zhu 2005; Woodward 2006). C) HR in turn results in an imbalance between RNR levels and levels of origin activation. Under these conditions DNA replication forks are expected to stall (lightning bolt), which, in association with unscheduled DNA replication, invokes a strong DDR (Chk1/Chk2) due to the amplified levels of ssDNA and double strand DNA breaks (Lupardus 2002; Di Micco 2007). Hence, following DNA damage, a positive feedback loop (HR) amplifies the negative feedback loop (DDR) that eventually leads to OIS. In the absence of positive feedback during a normal S phase, asynchronous activation of replication origins is a consequence of the checkpoint mediated balance between replication elongation and origin activation (black bar and red arrow in Figure 5).

fig.1

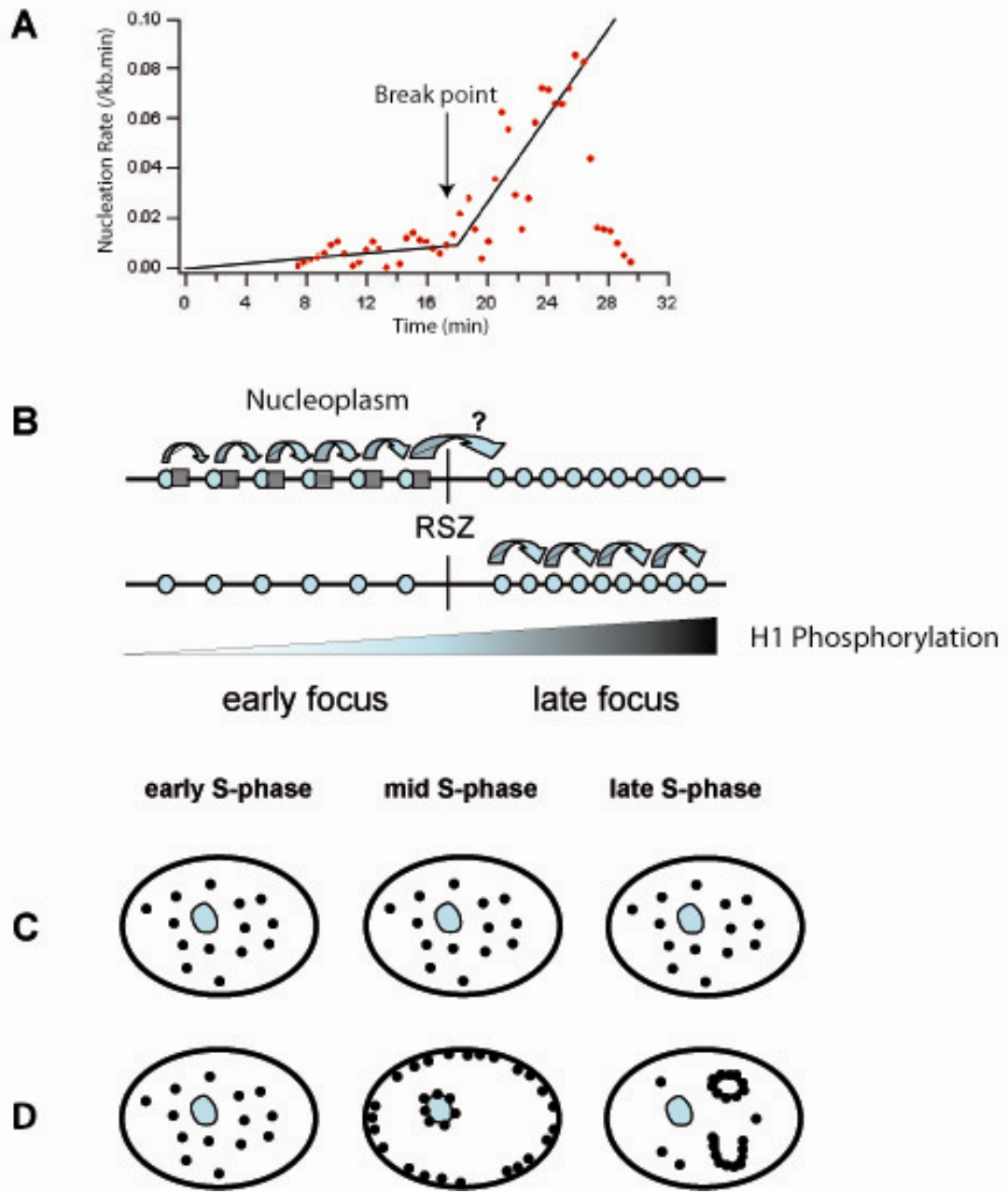

fig 2

A
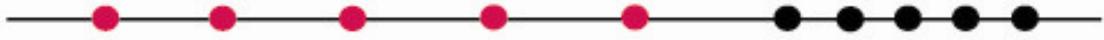
early    late    TDP

B
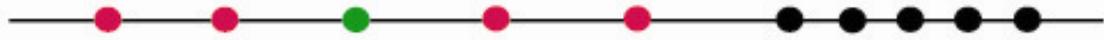

ODP

C
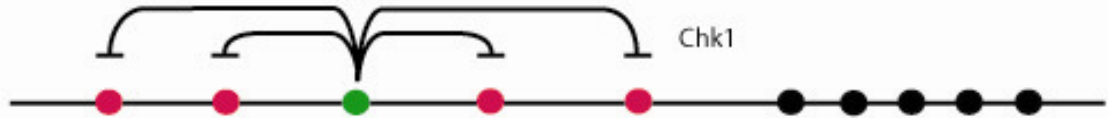

OI

D
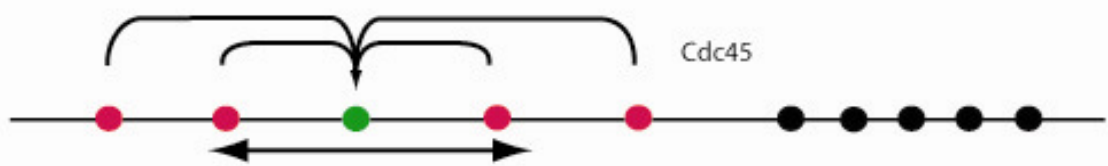

Initiation

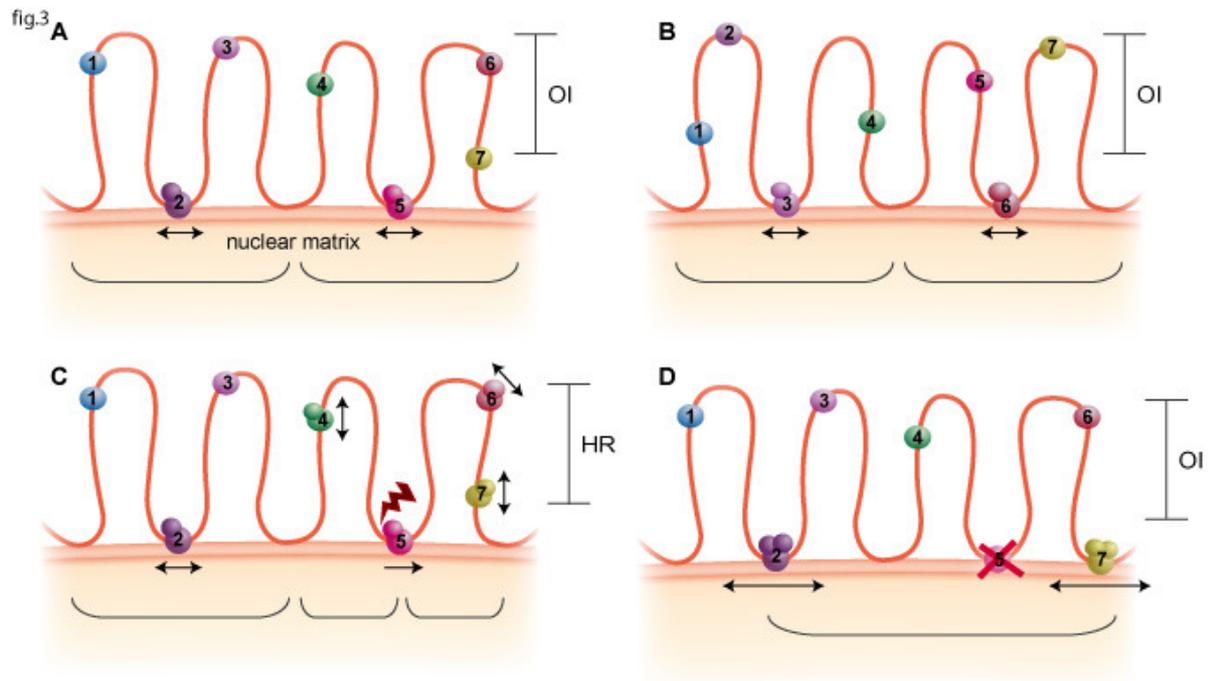

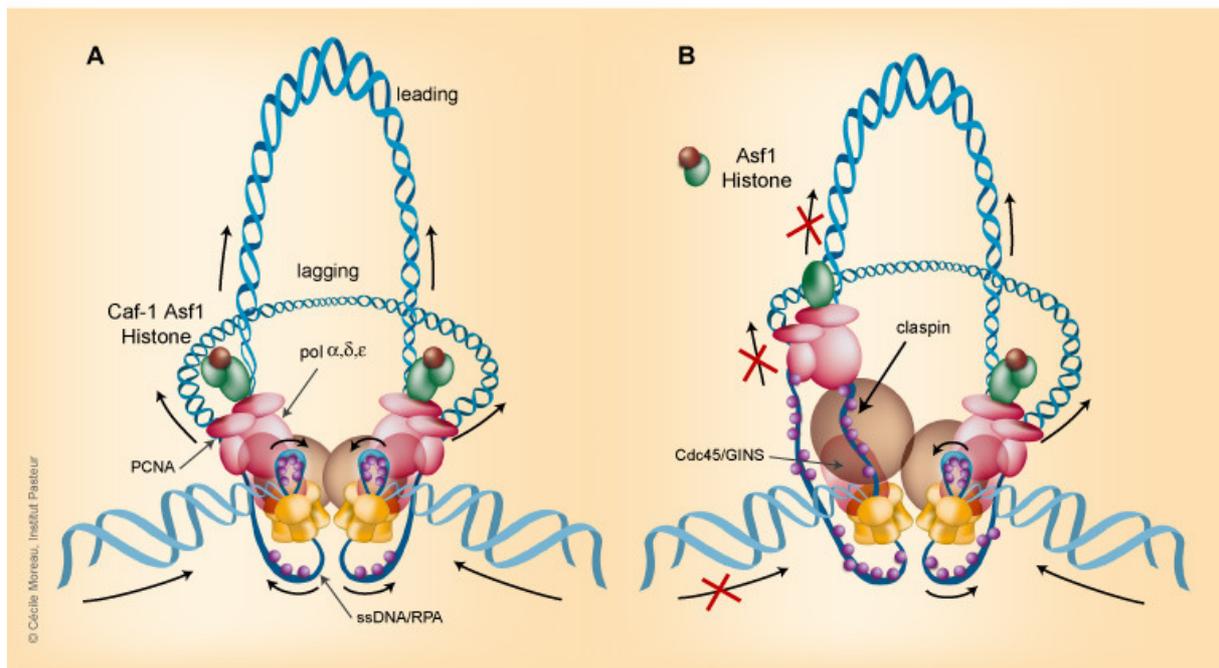

fig.5A

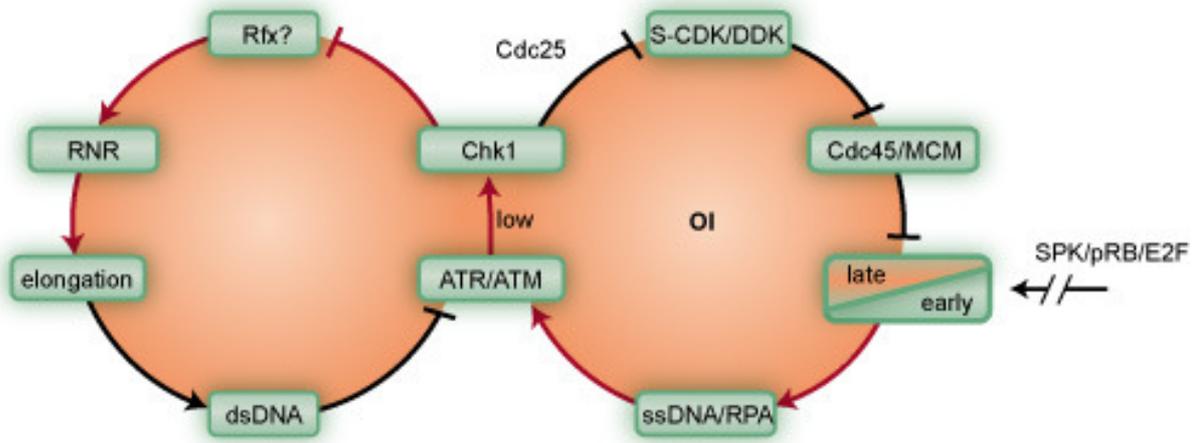

Fig 5B

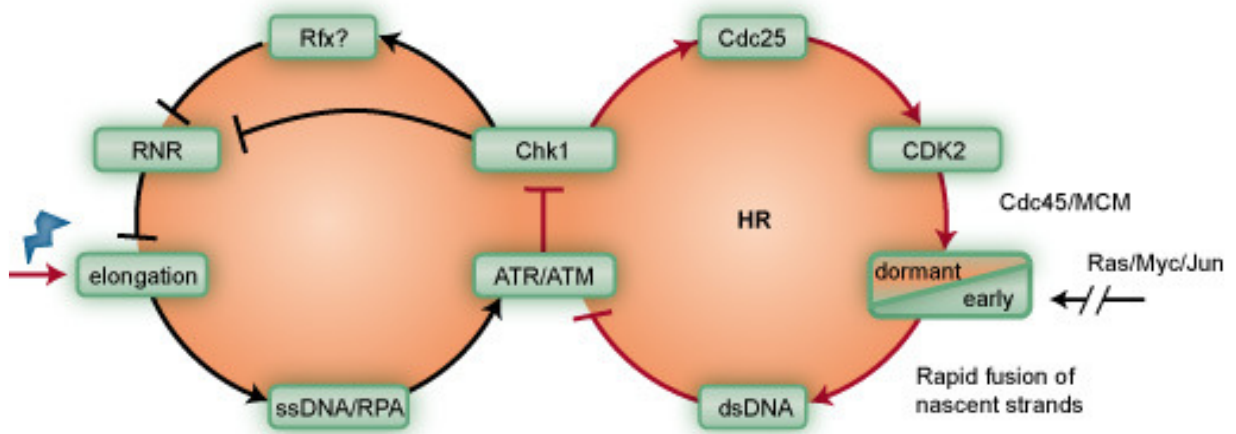

fig.5C

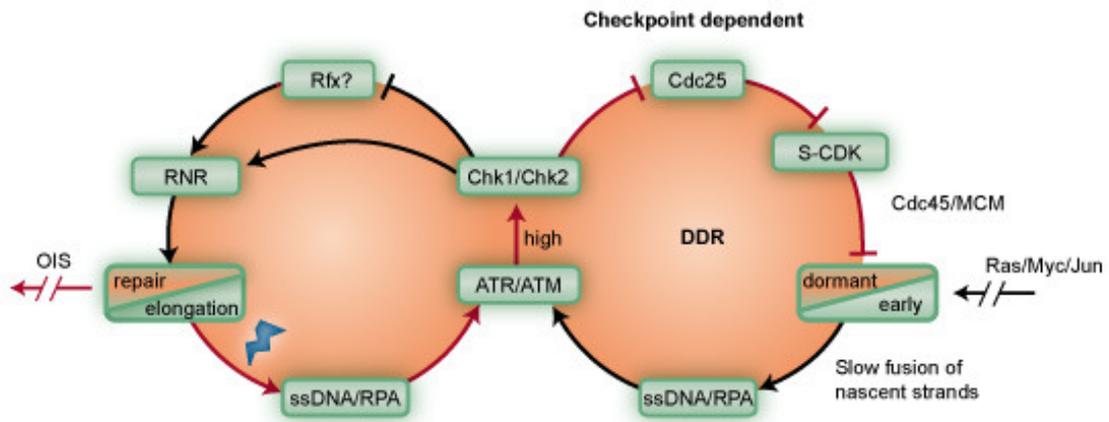